\documentclass[aip,reprint,amsmath,amssymb,twocolumn,floatfix,superscriptaddress]{revtex4-1}

\usepackage{graphicx}
\usepackage{hyperref}
\usepackage{siunitx}
\usepackage{amsmath}
\usepackage{placeins}

\newcommand{\ie}{\emph{i.\,e.}}
\newcommand{\eg}{\emph{e.\,g.}}
\newcommand{\Eth}{E_\mathrm{Th}}
\newcommand{\Reff}{R_\mathrm{eff}}

\begin{document}

\title{Modification of electron-phonon coupling by micromachining and suspension}

\author{Olli-Pentti Saira}
\email{osaira@bnl.gov}
\affiliation{Condensed Matter Physics and Kavli Nanoscience Institute, California Institute of Technology, Pasadena, CA 91125}
\affiliation{Computational Science Initiative, Brookhaven National Laboratory, Upton, NY 11973}
\author{Matthew H. Matheny}
\affiliation{Condensed Matter Physics and Kavli Nanoscience Institute, California Institute of Technology, Pasadena, CA 91125}
\author{Libin Wang}
\affiliation{QTF Centre of Excellence, Department of Applied Physics, Aalto University, FI-00076 Aalto, Finland}
\author{Jukka Pekola}
\affiliation{QTF Centre of Excellence, Department of Applied Physics, Aalto University, FI-00076 Aalto, Finland}
\author{Michael Roukes}
\affiliation{Condensed Matter Physics and Kavli Nanoscience Institute, California Institute of Technology, Pasadena, CA 91125}
\date{\today}

\begin{abstract}
Weak electron-phonon interaction in metals at low temperatures forms the basis of operation for cryogenic hot-electron bolometers and calorimeters. Standard power laws, describing the heat flow in the majority of experiments, have been identified and derived theoretically. However, a full picture encompassing experimentally relevant effects such as reduced dimensionality, material interfaces, and disorder is in its infancy, and has not been tested extensively. Here, we study the electron-phonon heat flow in a thin gold film on a SiO${}_2$ platform below 100~mK using supercurrent thermometry. We find the power law exponent to be modified from $5.1$ to $4.6$ as the platform is micromachined and released from its substrate. We attribute this change to a modified phonon spectrum. The findings are compared to past experiments and theoretical models. 
\end{abstract}

\maketitle

\section{Introduction}

Nanostructures at low temperatures exhibit strong thermal response. This follows from their minuscule heat capacities and thermal conductances, both of which diminish in response to the shrinking of the device dimensions and reduction of the operating temperature~\cite{Giazotto2006}. Consequently, low-temperature nanodevices excel as thermal detectors, \eg, bolometers and calorimeters, and various types of heat engines. Quantitative measurements of the thermal properties of materials and structures constitute a valuable tool for scientific and engineering purposes. For detector applications, thermal characterization is vital for understanding and optimizing the detector performance~\cite{Irwin2005}. Recently, in basic research, thermal transport has been used as a probe of otherwise elusive states of matter such as strongly interacting quantum Hall systems~\cite{Banerjee2017}, and the phonon spectrum in micromachined supports~\cite{Schwab2000}. Several thermal and thermoelectric signatures of Majorana states have been identified theoretically~\cite{Lopez2014,Ramos-Andrade2016,Molignini2017}.

Despite the above, studies of mesoscopic thermal transport are far outnumbered by studies of electrical transport. We speculate this is because the experimental configurations for quantitative thermal measurements tend to be complex. The associated technological challenges can be grouped into two categories. The first challenge is the measurement of the relevant physical quantities temperature, power, and energy. The measurement should be accurate (\ie, free of systematic biases), precise (\ie, capable of resolving small differences), and localized (\ie, only probe the targeted thermal body). A second challenge is ensuring the non-invasiveness of the measurement scheme. Attaching or operating the thermal probe may induce additional dissipation, or increase the thermal conductance or heat capacity of the target body, skewing the results. 

In this work, we advance the SNS weak link electronic supercurrent thermometry that has been developed and employed in several earlier studies~\cite{Dubos2001,Courtois2008,Govenius2014,Wang2018}. We show that our implementation of the method meets the criteria for ideal quantitative thermal characterization. We employ the method to study the nature of electron-phonon coupling in a mechanically suspended system that hosts a 3D electron gas and a quasi-2D phonon system. Our main finding is a modified electron-phonon heat flow $\dot{Q} \propto T_{el}^n - T_{ph}^n$ with $n \approx 4.6$, where $T_{el}$ and $T_{ph}$ denote the electron and phonon temperatures, respectively. Very recent theoretical analysis~\cite{Anghel2018} of the metal-dielectric bilayer system has predicted a "plateau region" where $4.5 < n < 5$ over a wide range of temperatures where the phonon system undergoes a 2D to 3D transition. Our work appears to confirm this prediction. An exponent $n \approx 4.5$ has also been seen in an earlier experimental study that employed different materials, sample geometry, and thermometry method from ours~\cite{Karvonen2007,Karvonen2007b}.

\section{Methods}

\subsection{Physics of supercurrent thermometry}

Any physical phenomenon with a temperature dependence can be used as a thermometer after calibration against a known thermometer. This is called secondary thermometry. Here, we employ proximity superconductivity induced in a metallic SNS (superconductor--normal--superconductor) weak link~\cite{Shepherd1972,Clarke1969}. The switching current offers a virtually ideal electrical characteristic for steady-state thermometry using a simple readout circuit. The temperature range of the sensor is limited, but different temperatures can be targeted by choosing an appropriate length for the weak link.

\begin{figure}[t]
    \includegraphics[width=3.375in]{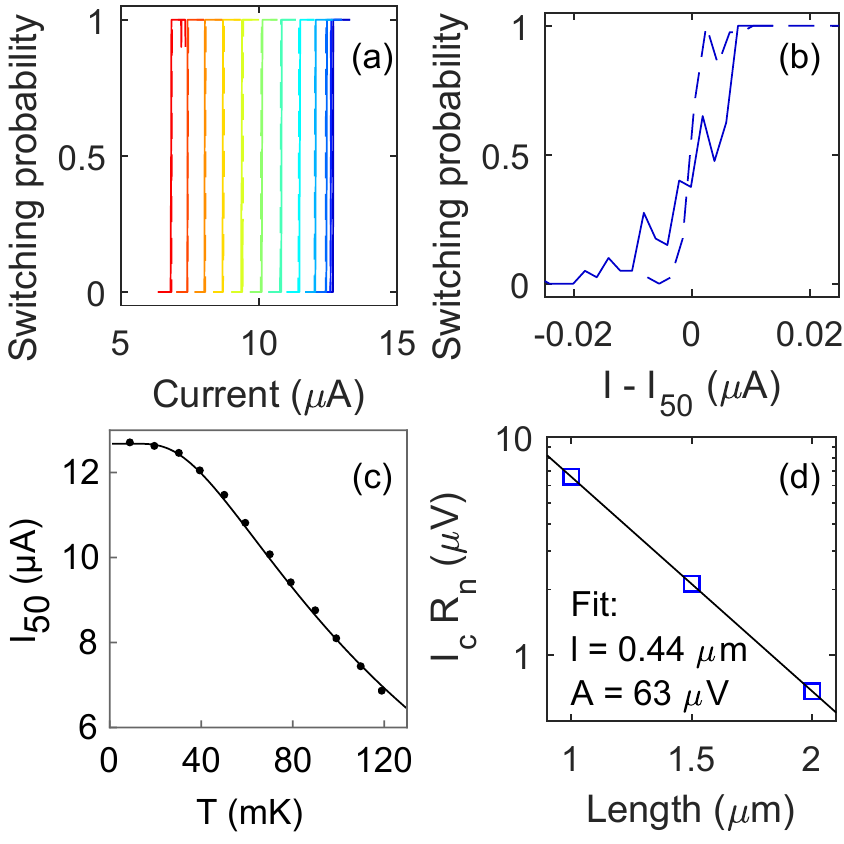}
    \caption{Basic characterization of diffusive SNS links. (a) Switching probability histograms at different temperatures for an $l = \SI{1.5}{{\mu}m}$ weak link. (b) A high-resolution scan of histogram at base temperature for both positive (solid) and negative (dashed) polarities. (c) The mean switching current $I_{50}$ extracted from the histograms in panel (a) as a function of temperature (markers), and the theoretical critical current using fitted parameter values (line). (d) Measured base-temperature $I_c R_N$ product for three different wire lengths, and an exponential fit.}
    \label{fig:SNS}
\end{figure}

The electrical behavior of the SNS weak link can be analyzed in terms of the RCSJ model~\cite{Stewart1968,McCumber1968}. When probed with a low-frequency current waveform, the junction stays essentially in a zero-voltage state until the probe current exceeds the temperature-dependent critical current of the junction. Crucially, the dissipation in the zero-voltage state is vanishingly small. After the junction enters a finite-voltage state, positive electro-thermal feedback brought upon by the current bias rapidly heats up the junction to a temperature where $I_c = 0$ and $V = R_n I$.

A basic characterization of an underdamped weak link can be performed by determining its switching current histogram. We fix the shape and duration of a current waveform, and determine the probability of observing a voltage pulse as a function of the amplitude. In Fig.~1(a), we have used a single cycle of a 100~ms period sinewave to probe an SNS weak link with an approximate separation of 1.5~$\mu$m between the superconducting electrodes at temperature ranging from 10~mK to 120~mK. The sample geometry is described in more detail in Sec.~\ref{sec:bulk}. To quantify the width of the histograms [Fig.~1(b)], we evaluate the standard deviation of the switching current, which is found to be less than 16~nA at all temperatures.

To quantify the effect of temperature on the position of the switching histogram, we evaluate the mean switching current at each temperature point [Fig.~1(c), markers]. The temperature dependence obtained in this manner can be reproduced by a low-temperature expansion of the supercurrent of a diffusive SNS weak link~\cite{Likharev1979,Dubos2001}
\begin{eqnarray}
    \Eth & = & \hbar D/L^2\nonumber\\
    e \Reff I_c / \Eth & = & a \left(1 - b e^{-a \Eth/(3.2 k_B T)}\right)
\end{eqnarray}
where $\Eth$ denotes the Thouless energy, $D$ is the diffusion constant, $L$ is the physical length of the weak link, $R_{eff}$ is the normal-state resistiance, $I_c$ is the critical current, $T$ is the temperature, and $a = 10.82$, $b = 1.3$ are numerical constants. A separate four-wire characterization yields $D = \SI{240}{{cm}^2 s^{-1}}$ for the Au film, and $R_N = \SI{0.27}{\Omega}$ for this particular weak link. The other parameters appearing in the theory can be fitted to be $L=\SI{2.2}{{\mu}m}$ and $\Reff = \SI{2.7}{\Omega}$ [Fig.~1(c), line]. In a rectangular wire geometry, the ratio $R_N / \Reff \approx 0.1$ could be interpreted as a measure of the interface transparency~\cite{Courtois2008,Jabdaraghi2016}. However, in the more complex geometry employed here, there is additional normal metal that shunts the supercurrent link, thus lowering $R_N$ but not contributing to $I_c$.

The single-shot temperature resolution of the detector can be calculated by dividing the width of the histogram by the local temperature responsivity $\left|dI_{mean}/dT\right|$, which peaks at  $\SI{70}{nA/mK}$ at $T = \SI{70}{mK}$. The detector is capable of resolving temperature differences of the order of 0.2~mK in its most sensitive range of temperatures with a single readout pulse.

Another characteristic prediction of the diffusive theory is the strong length dependence of the induced superconductivity. We investigate this by determining the base temperature $I_c R_N$ product for SNS wires of 500~nm width and varying length [Fig.~1(d)]. The data is consistent with an exponential decay with a characteristic length of 440~nm.

The position and width of the histogram can be weakly affected by factors such as the probing waveform, current and voltage noise, and the decision threshold for the detection of voltage pulses. However, as long as the same excitation and detection methods are used throughout the experiment, secondary thermometry incurs no systematic bias. Two-level composite pulses have been used in previous works to greatly improve the temporal accuracy of the supercurrent probing~\cite{Zgirski2018}. This work deals exclusively with steady-state quasi-equilibrium thermometry, thereby obviating the need for more complex pulse sequences.

Since a hysteretic Josephson junction acts as a wide-bandwidth threshold detector for current noise, we estimate the magnitude of temperature-dependent current noise that reaches that junction. The twisted-pair measurement lines had two RC-filter banks with $R = \SI{600}{\Omega}$ and $\left(RC\right)^{-1} = 2\pi\times\SI{100}{kHz}$ each at the 4K and base temperature stage of the fridge. Hence, Johnson-Nyquist noise from outside of the mixing chamber stage should be negligible. Between the final RC filter and the sample, there is a powder filter segment consisting of 20~cm of resistive constantan twisted pair wire ($R \approx \SI{14}{\Omega}$ per wire) embedded in a lossy Stycast/Cu powder dielectric. Based on room-temperature characterization of a similar filter box, we estimate the powder filter segment to strongly attenuate signals above $f_c = \SI{100}{MHz}$. We then estimate the worst-case rms current amplitude of noise from the sample box wiring reaching the Josephson junction to be $I_{rms} = \sqrt{4 k_B T f_c/(2R)} = \SI{5}{nA}$, where we substituted $T = \SI{0.12}{K}$ corresponding to the maximum temperature encountered in this work.

\subsection{Heat conductance measurements}

\begin{figure}[t]
    \includegraphics[width=3.375in]{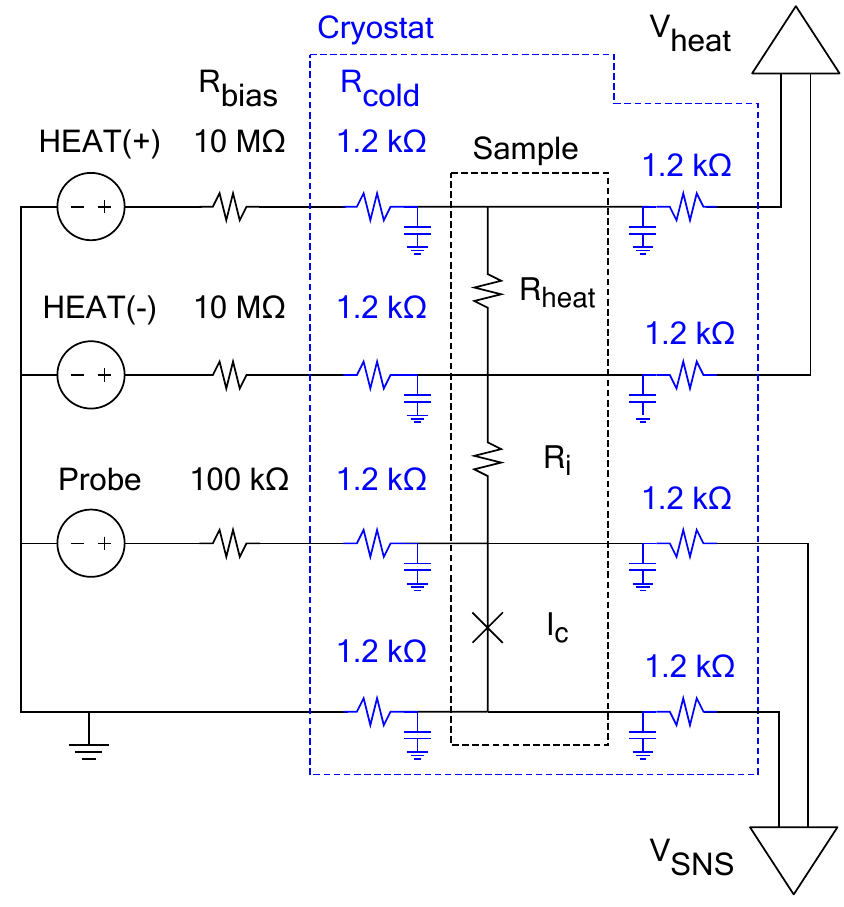}
    \caption{Electrical schematic for measurements reported in this letter. The output voltage of the HEAT$(-)$ source is always the negative of output voltage of the HEAT$(+)$ source. The amplifiers are high input-impedance ($R_{in} > \SI{1000}{M\Omega}$) differential voltage preamplifiers.}
    \label{fig:Electrical}
\end{figure}

The subject of study in steady-state thermal measurements is the heat flow between two thermal bodies in quasi-equilibrium, \ie, assuming they both have well-defined internal temperatures. We denote the bodies by $a$, $b$ and their temperatures by $T_a$ and $T_b$ from here on.  For elementary, continuous systems of interest to mesoscopic physics, the heat flow can be generally written in the form
\begin{equation}
    \dot{Q}_{a \rightarrow b} = h(T_a)-h(T_b),\nonumber
\end{equation}
where $h$ is a monotonically increasing function. In a limited temperature range, one often finds a \emph{power law}
\begin{equation}
    \dot{Q}_{a \rightarrow b} = A (T_a^n - T_b^n),\label{eq:Qdot}
\end{equation}
where the exponent $n$ often conveys information about the dimensionality of the microscopic physics of energy transport. Equivalently, one can study the thermal conductance
\begin{equation}
    G(T) = \left.\frac{\partial \dot{Q}_{a \rightarrow b}}{\partial T_a}\right\vert_{T_a = T} = h'(T_a) = n A T^{n-1},
\end{equation}
where the last equality holds for any power law. Throughout this manuscript, we will use exponent $n$ to refer to the heat flow power law [Eq.~(\ref{eq:Qdot})]. Consequently, the exponent for the thermal conductance will be $n-1$.

Accurate experimental investigations of the above relations require independent determination of the three quantities $\dot{Q}_{a \rightarrow b}$, $T_a$, and $T_b$. In some cases, one can substitute in place of $T_b$ the reading of another thermometer $T_c$. This is possible if the thermal link between $b$ and $c$ is strong enough to disallow significant temperature differences $\left|T_b - T_c\right|$, or if a non-zero direct thermal link between $b$ and $c$ exists, and $\dot{Q}_{b \rightarrow c} = 0$ is known to sufficient accuracy.

\begin{figure}[t]
    \includegraphics[width=3.375in]{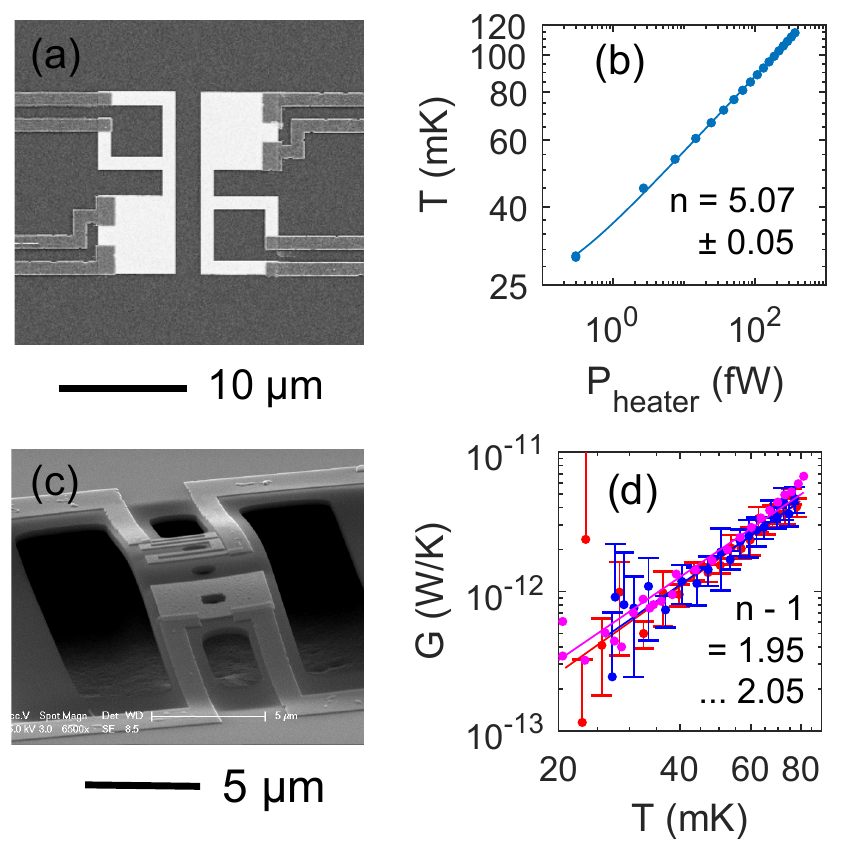}
    \caption{Verification experiments. \emph{Electron-phonon coupling on bulk substrate:} (a) SEM image of a device nominally identical to the measured one. (b) Electron temperature as a function of heating power and a power law fit. \emph{Phononic thermal conductance of a micromachined structure:} (c) SEM image of the measured device. (d) Three thermal conductance datasets. Data from the small membrane [device in panel (c), magenta markers], and the large membrane [device in Fig.~\ref{fig:Platform2}(b), red and blue markers corresponding to two independent measurements]. Thermal conductance obtained as the local numerical derivative of the measured heating characteristic. Lines are power-law fits to the data with a background heating offset.}
    \label{fig:Verification}
\end{figure}

We will discuss how supercurrent thermometry allows one to approach the ideal thermal measurement in practice with a relatively simple measurement setup. The electrical connections for a single-body electron thermometry experiment are illustrated in Fig.~\ref{fig:Electrical}. The metallization pattern on the chip, corresponding to the "Sample" sub-circuit in the diagram, can be seen in Fig.~\ref{fig:Verification}(a). Our metallic samples containing no tunnel junctions present low electrical impedances to the measurement circuitry. Referring to the labels of Fig.~\ref{fig:Electrical}, the largest resistance on the chip is the heater wire, $R_{heat} \approx \SI{10}{\Omega}$. For direct electron heating and thermometry, a galvanic connection between the heater and thermometer is required. The electrical resistance $R_i$ of this connection is less than $\SI{1}{\Omega}$. In the following, we show that the amount of current flowing through this connection is negligible.

Due to the low impedance of the sample, it is straightforward to realize an accurate low-frequency current bias using external voltage sources and large bias resistors. Operating the voltage sources in such a manner that $V_{heat(\pm)} = \pm V_{heat}$ ensures that the heating current $I_{heat} = V_{heat}/R_{bias,heat}$ flows through the resistor $R_{heat}$. To the precision within which equal and opposite current biasing is realized, \emph{no} current flows through $R_i$ or the Josephson junction (JJ) modeling the SNS weak link. 
Similarly, the probe source biases the Josephson junction with the current $I_{probe} = V_{probe}/R_{bias,probe}$, and only a small fraction of it flows through $R_i$ or $R_{heat}$. More quantitatively, the relative leakage through $R_i$ is $2R_{cold}/R_{bias,heat} \approx 2\times 10^{-3}$, and the leakage through $R_{heat}$ is half of that.

We can verify the accuracy of the current-cancellation scheme in situ: If a fraction $\alpha$ of $I_{heat}$ were to flow through $R_i$ and the JJ element, one would observe a switching current $I_c \pm \alpha I_{heat}$ that depends on the polarity of the heating current. We find $|\alpha| < 1\%$. Consecutively, the maximum leakage current encountered in the experiment is less than 2~nA, contributing to a maximum of $\SI{4d-18}{W}$ of unaccounted heating. To further null the effect of the heating current leakage, and any other unaccounted current offsets, we always test all four combinations of polarities for the bias and the heating currents.

The voltage drop $V_{heat}$ over the heating wire can be accurately measured with a high-input impedance voltage preamplifier. The heating power in the target electron gas can be then evaluated as $P = I_{heat} V_{heat}$. Since the heating wire is a resistive element with a linear, temperature-independent $I$-$V$ characteristic, we forego the voltage measurement after the resistance $R_{heat}$ has been determined and use $P = R_{heat} I_{heat}^2$ instead. Finally, we note that our use of leads made of only superconducting materials ensures that all on-chip dissipation takes place in the normal metal reservoirs, and that electronic heat conductance along the leads is suppressed to a negligible level. Since the JJ used for temperature measurement stays in zero-voltage state until an output voltage is generated, there is no spurious heating from the operation of the thermometer that would need to be included in the model.

\subsection{Sample fabrication}

The devices for this study were fabricated on a lightly p-doped (resistivity 10--100 $\Omega$\,cm) silicon substrate with 300~nm of thermally grown SiO${}_2$ on its surface. The normal-state (3~nm Ti, 50~nm Au) and superconducting (5~nm Ti, 65~nm Al) films were deposited separately using e-beam lithography, e-beam evaporation, and lift-off processes. In the normal-state film, the superconducting character of the Ti adhesion layer can be expected to be completely suppressed by the inverse proximity effect. The surface of the Au film was cleaned with in-situ Ar ion milling before the superconducting contacts were formed. To create the mechanically suspended structures, the SiO${}_2$ film was first patterned using e-beam lithography and Ar ion milling. Finally, the SiO${}_2$ platform and the metallic structures were released by removing the underlying Si with an isotropic XeF${}_2$ etch.

\section{Thermal measurements}

\subsection{Electron-phonon coupling in bulk}
\label{sec:bulk}


Early work on measurement of current-induced disequilibrium between electrons and phonons was carried out using local noise thermometry~\cite{Roukes1985}. We realize a similar non-suspended device geometry with heater and thermometer connections [Fig.~3(a), only one half of the device was tested]. We find $n = 5.07\ (\sigma = 0.03)$; $A = 2.23 \times 10^{-8}\, \mathrm{W K^{-n}}\ (\sigma_{rel} = 0.01)$ [Fig.~3(b)]. The fitted exponent is very close to the value $n = 5$ corresponding to a clean-limit 3D metal~\cite{Wellstood1994}. We note that the calculation presented in Ref.~\onlinecite{Qu2005} predicts a slightly higher temperature-dependent exponent for a thin metallic film on a bulk dielectric. The nominal volume of the gold island is $V = \SI{57}{{\mu}m^2} \times \SI{50}{nm} = \SI{2.9}{{\mu}m^3}$. The fitted $A$ yields an electron-phonon coupling constant $\Sigma = A / V = \SI{7.7d9}{W K^{-n} m^{-3}}$. We have neglected the effect of the Ti adhesion layer, which makes up 4\% of the thickness of the normal-state film. The only previously reported measurement of $\Sigma$ for Au that we are aware of is $2.2 \ldots \SI{3.3d9}{W K^{-5} m^{-3}}$ from Ref.~\onlinecite{Echternach1992}, which employed a much thinner and more disordered film ($d = \SI{11.2}{nm}$, mean free path $\SI{2.5}{nm}$) compared to ours.

\subsection{2D Phononic thermal conductance}
\label{sec:ph}




We have studied two different membrane designs. The smaller membrane [Fig.~3(c)] has four support legs with a nominal width of $w = \SI{2}{{\mu}m}$ and length $l = \SI{7.5}{{\mu}m}$. The geometry factor is $4w/l = 1.07$. We find $n  = 2.98\ (\sigma = 0.03)$; $A = 2.46\times10^{-10}\, \mathrm{W K^{-n}}\ (\sigma_{rel} = 0.02)$ [magenta data in Fig.~3(d)].

The larger membrane [Fig.4~(b)] has four support legs with a nominal width of $\SI{4.5}{{\mu}m}$ and length $\SI{10.25}{{\mu}m}$. We find $n  = 3.05\ (\sigma = 0.02)$; $A  = 2.62\times 10^{-10}\, \mathrm{W K^{-n}}\ (\sigma_{rel} = 0.01)$ using the Right heater/Left thermometer combination [red data in Fig~3(d)], and $n = 3.01\ (\sigma = 0.02)$; $A = 2.37\times 10^{-10}\, \mathrm{W K^{-n}}\ (\sigma_{rel} = 0.01)$ using the Left heater/Right thermometer combination [blue data in Fig~3(d)]. 


The fitted exponents are very close to the integer value $n=3$ (corresponding to $\kappa \propto T^2$). This would indicate that the phonons are quantized in the thickness direction, but not in the lateral direction. The prefactor $A$ is appears to be independent of the geometry factor, indicating that phonon scattering in the support legs is weak. Phononic thermal conductance in suspended and patterned structures with a similar geometry have been analyzed theoretically in detail in Refs.~\onlinecite{Anghel1998,Kuhn2004}.

\subsection{Electron-phonon coupling in a thin suspended system}

\begin{figure}[t]
    \includegraphics[width=3.375in]{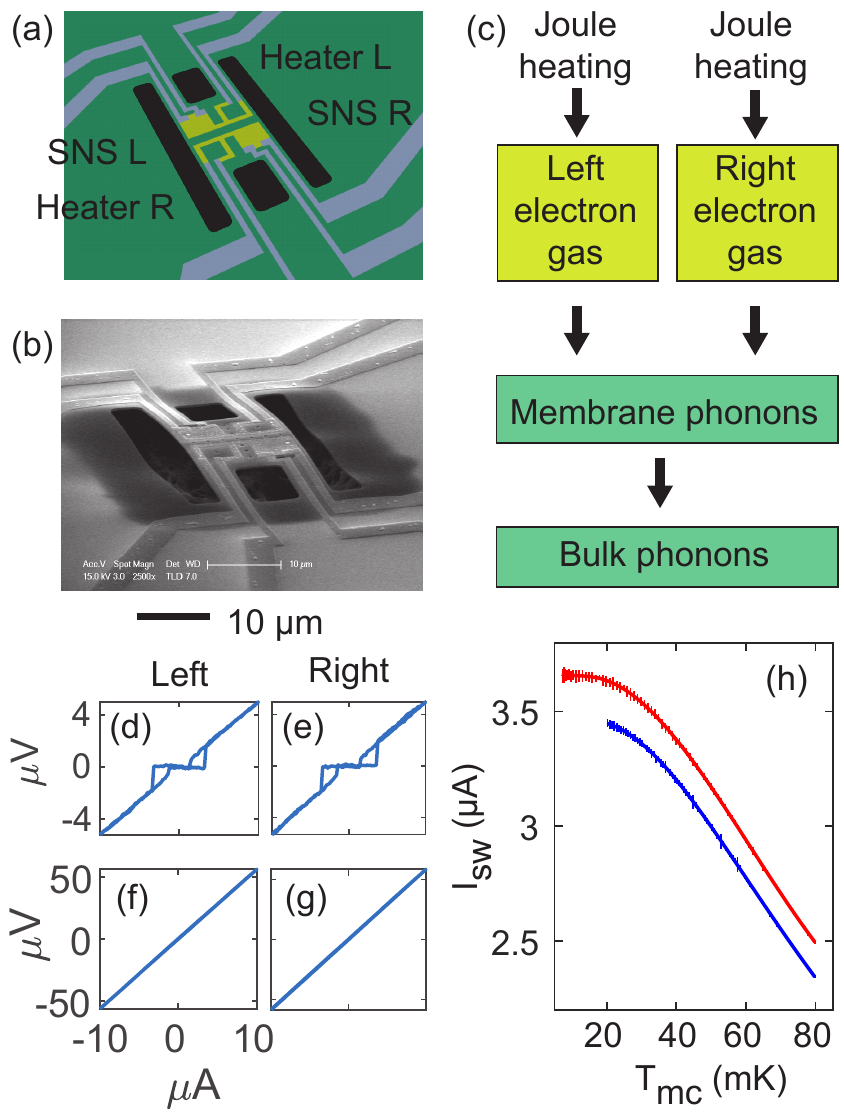}
    \caption{Experiment for studing electron-phonon coupling on suspended membrane. (a) Annotated 3D rendering of sample. (b) SEM image of the measured device. (c) Thermal block diagram of the experiment. (d)-(f) Base temperature I-V characteristics of the thermometers and heating elements on the platform. (h) Observed temperature dependence of the two critical current thermometers: Left (blue markers), Right (red). The thickness of the line indicates the standard deviation (2$\sigma$) evaluated from 30 measurements per temperature point.}
    \label{fig:Platform2}
\end{figure}

\begin{figure}[t]
     \includegraphics[width=3.375in]{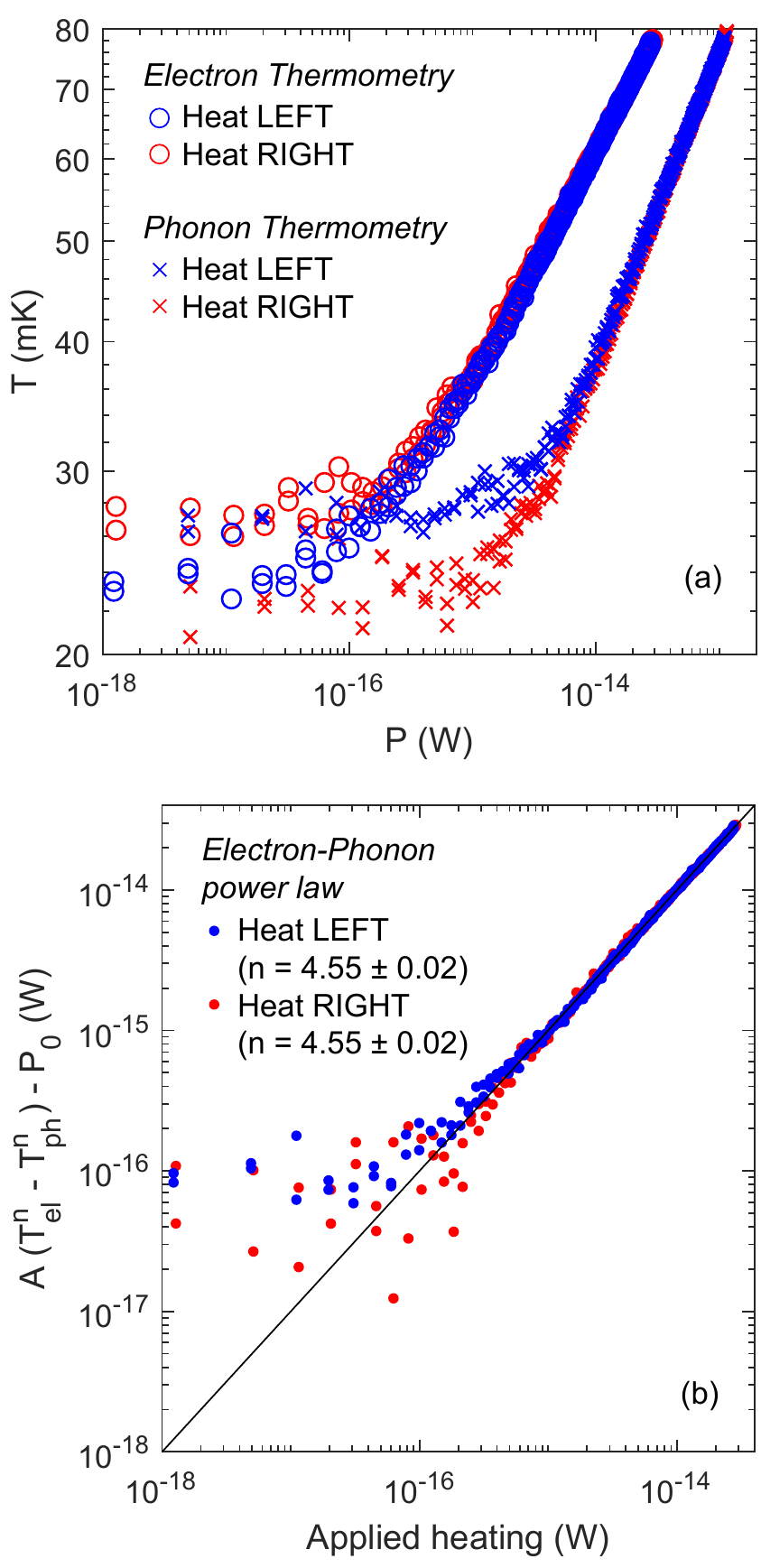}
    \caption{Electron-phonon power law. (a) The observed heating characteristic at base temperature for all four heater and thermometer combinations. (c) Comparison of the applied heating power  and the calculated electron-power heat flow using the power law in Eq.~(\ref{eq:eph_main}) with best-fit parameters.}
    \label{fig:EPhLaw}
\end{figure}

%


To study the electron-phonon coupling on a suspended membrane, we have fabricated a device that contains two symmetric heater-thermometer structures. Figure 4(a) shows a schematic of the device with labeled components; Fig.~4(b) is a scanning electron micrograph of the completed device. The upward buckling of the platform is caused by stresses in SiO$_2$ layer from the film growth process (see Appendix~\ref{sec:app:FEM}). The thermal block diagram that forms the basis of the quantitative analysis is shown in Fig.~4(c). Base temperature $I$-$V$ characteristics of the four electrical components of the device are shown in Figs.~4(d)-(g). The temperature calibration curves of the two supercurrent thermometers are shown in Figs.~4(h).

The critical currents after suspending the membrane are a factor 3--4 smaller than in the sample with bulk dielectrics. We suspect this is caused by degradation of the SNS weak link due to the additional processing steps and ageing of the sample. While the principle of supercurrent thermometry is still applicable, the temperature range where digital switching dynamics are observed is restricted to below 80~mK.

We wish to study the electron-phonon heat flow that we postulate to follow a power law
\begin{equation}
    \dot{Q} = A (T_{el}^n - T_{ph}^n) - P_0,\label{eq:eph_main}
\end{equation}
where the excess heating term $P_0$ accounts for the background heat loads that are beyond experimental control and are responsible for saturation of the thermometer signal at low bath temperatures and low heating powers. Details of the thermometer calibration and data analysis are presented in Appendix~\ref{sec:app:thermometry}. The dual-heater, dual-thermometer configuration allows us to independently measure the quantities $\dot{Q}$, $T_{el}$, and $T_{ph}$ appearing in the above equation. By swapping the roles of the left and right halves of the nominally symmetric design, we obtain two independent measurements of the power law.  We find $n = 4.55\ (\sigma = 0.01)$; $A = 3.78\times10^{-9}\, \mathrm{W K^{-n}}\ (\sigma_{rel} = 0.003)$ using Right heater data [Red markers in Figs.~5(a), (b)], and $n = 4.55\ (\sigma = 0.01)$; $A = 3.75\times10^{-9}\, \mathrm{W K^{-n}}\ (\sigma_{rel} = 0.003)$ using Left heater data [Blue markers in Figs.~5(a), (b)]. The two independent measurements of the exponent and the magnitude of the electron-phonon heat flow are remarkably close to each other. The exponent is in agreement with earlier studies by Karvonen and Maasilta~\cite{Karvonen2007,Karvonen2007b}, where they obtained $n \approx 4.5$ at temperatures up to $\SI{0.4}{K}$ employing quasiparticle thermometry on a SiN membrane. The fitted value for the excess heating term is $\SI{0.9e-16}{W}$ and $\SI{-2.1e-16}{W}$ for the Right and Left heater data respectively. A negative $P_0$ indicates that the background heating was higher in the phonon temperature measurement configuration compared to the electron temperature measurement.

The theoretical work that most closely models our experimental system is that by Anghel and coworkers~\cite{Cojocaru2016,Anghel2017,Anghel2018}, wherein they consider a sheet consisting of finite-thickness metal and dielectric layers. In Ref.~\onlinecite{Anghel2018} in particular, they extend their analysis to cover the range of temperatures where the phonon dimensionality changes from two to three. A key parameter of their model is the crossover temperature, for which they give the homogeneous media approximation $T_C \approx \hbar c_t /(2 k_B d)$, where $c_t$ is the transversal (shear wave) speed of sound, and $d$ is the total thickness of the membrane. We estimate $T_C$ for our device as $$T_C \approx \frac{\hbar}{2k_B} \left(\frac{d_{\mathrm{SiO2}}}{c_{t,\mathrm{SiO2}}} + \frac{d_{\mathrm{Ti}}}{c_{t,\mathrm{Ti}}} + \frac{d_{\mathrm{Au}}}{c_{t,\mathrm{Au}}}\right)^{-1} = \SI{30}{mK}.$$ Hence, our experiment covers the range $T/T_C = 0.7 \ldots 2.7$. In the numerical calculation of Ref.~\onlinecite{Anghel2018}, the local power-law exponent $\frac{\partial \ln  \dot{Q}}{\partial\ln T_{el}}$ is found to lie between 4.5 and 4.7 for $T/T_C = 2 \ldots 10$, constituting the plateau region.



The quality of our experimental data does not permit a reliable extraction of local exponents. However, the global exponent given by an unweighted fit is dominated by the data at higher temperatures, and therefore measures primarily the plateau exponent. To verify this, we have performed a separate power law fit using only the data where $T_{el} \geq 2T_C = \SI{59}{mK}$, which yields $n = 4.56\ (\sigma = 0.02)$ using Right heater data, and $n = 4.51\ (\sigma = 0.02)$ using Left heater data.

In conclusion, we have used an SNS critical current thermometer to study electron-phonon coupling in a micromechanical platform. Comparing the measurements before and after micromachining and suspension of the platform, we find the electron-phonon power law to be affected by the change in the local phonon spectrum.  In particular, we have observed an exponent $4.51 \leq n \leq 4.56$ in the suspended membrane in the plateau temperature region corresponding to the phononic 2D-to-3D crossover, in agreement with a recent theoretical prediction.

\emph{Acknowledgments} We thank D.-V. Anghel for important discussions during the preparation of the manuscript. This material is based upon work supported by, or in part by, the U.~S. Army Research Laboratory and the U.~S. Army Research Office under contracts W911NF-13-1-0390 and W911NF-18-1-0028, and Academy of Finland under contract 312057.

\appendix

\section{Simulation of platform deformation}
\label{sec:app:FEM}

\begin{figure}
    \includegraphics[trim={0.5in 0.3in 0.5in 0.4in},clip,width=3.375in]{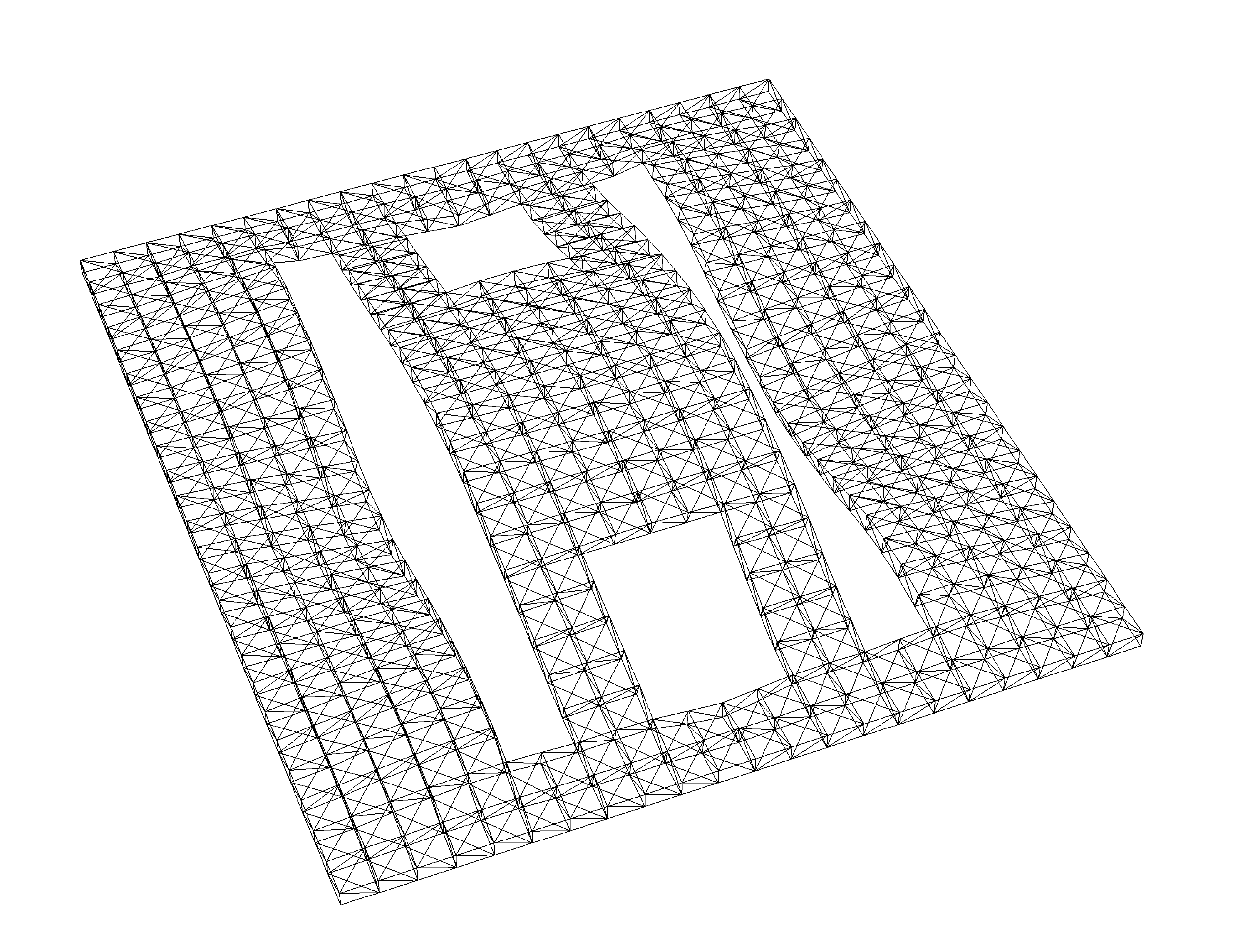}
    \caption{Simulated deformation of the approximate large membrane geometry under isotropic strain $\epsilon = -0.029$ applied to the support frame. The lateral size of the mesh is approximately $\SI{2}{{\mu}m}$. To be compared with Fig.~\ref{fig:Platform2}(b).}
    \label{fig:app:FEM}
\end{figure}

We estimate the residual stress in the films, as it has the potential to affect the electronic and phononic structure, and hence the thermal properties. A literature value for thermally grown SiO${}_2$ is $\SI{2d10}{dyn\ cm^{-2}}$ compressive~\cite{Lin1972}, and for evaporated metals one finds $2\ldots\SI{9d9}{dyn\ cm^{-2}}$ tensile~\cite{Klokholm1969}. The stress forces become evident in the deformation of the structure after the underlying Si substrate is removed. Ignoring the effect of the metal films, and using $E = \SI{70}{GPa}$ as the Young's modulus of SiO${}_2$, we predict an engineering strain $\epsilon = -0.029$. The observed distortion is well reproduced by a finite-element model where the non-suspended surroundings of the membrane are rigidly compressed by that amount. The result of the finite-element analysis is shown in Fig.~\ref{fig:app:FEM}.

At the cryogenic temperatures of the experiment, the stress and strain will be different due to thermal contraction. We did not possess the capability to directly measure the deformation of the membrane  in the cryostat. Instead, we note the literature values of the total linear thermal contraction $(l_{\SI{0}{K}}-l_{\SI{300}{K}})/l_{\SI{300}{K}}$ of the constituent materials~\cite{Corruccini1961}: $\SI{-22d-5}{}$ for Si; $\SI{8d-5}{}$ for SiO${}_2$; $\SI{-324d-5}{}$ for Au; and $\SI{-415d-5}{}$ for Al. All of these are at least an order of magnitude smaller than the strain induced by the intrinsic stress in the SiO${}_2$ film and therefore will not significantly alter the deformation observed in the SEM micrograph.

\section{Secondary thermometry}
\label{sec:app:thermometry}
In this section, we consider possible systematic errors that arise in the measurements of thermal properties of nanostructures performed with secondary thermometers. By definition, a primary thermometer is one that does not need calibration against other thermometers. Primary thermometers are based on a physical measurands with an \emph{a priori} known temperature dependence.  A secondary thermometer can be constructed from any temperature-dependent measurand by calibrating it against a known thermometer. In this work, we use extensively secondary, local electron thermometry that is based on the critical current $I_c$ of an SNS weak link.

In principle, it is possible to calculate the $I_c(T)$ dependence from physical properties of the weak link. In practice, however, one cannot independently and accurately determine all factors affecting the measured critical current. Instead, one calibrates the empirical $I_c$ against the cryostat thermometer, which is assumed to be traceable to primary temperature standards. Many practical thermometers are nonlinear and suffer from loss of sensitivity at either end of their usable temperature range. Both of these apply to the critical current thermometer. For the following analysis, we only assume the existence of a measurand $M(T)$ that is a monotonic function of the temperature $T$ of the mesoscopic system of interest.

Virtually all low-temperature mesoscopic systems are susceptible to background heating. We use the term background heating to refer to an inflow or energy to the system under study that is not under the control of the experimentalist. For bolometer-like electronic systems, such as ours, a common source of background heating is electronic noise that originates from hotter temperature stages of the refrigerator, or from room temperature electronics. Although filtering and shielding of the sample and experimental wiring helps, the heating typically cannot be completely eliminated.

The question we wish to address here is whether systematic errors can be avoided in a measurement that employs a non-linear secondary thermometer, and is subject to non-zero background heating.

\subsection{Single thermal body}
We denote the temperature of the mesoscopic electron system by $T_e$. We assume that the experimentalist has control over the cryostat temperature $T_b$, and can apply an additional heating $P_{ext}$ to the electron system. We will assume that the background heating is independent of both $T_e$ and $T_b$, which is reasonable for external noise sources, and has magnitude $P_0$. We model the experiment under steady state heating as
\begin{eqnarray}
    P_{ext} + P_0 & = & h(T_e) - h(T_b)\label{eq:e}\\
    M             & = & M(T_e),
\end{eqnarray}
where the unknown function $h$ describes the heat flow between the electron system and the surrounding thermal bath. For a 3D electron system cooled only by electron-phonon coupling, one would have $h(T) = \Sigma V T^5$. The thermometer measurand $M$ is a function of $T_e$ only.

In the first step of the experiment, one performs a calibration of $M(T_e)$ by setting $P_{ext} = 0$ and sweeping $T_b$. In all physically relevant scenarios, $h(T)$ grows at least quadratically (superlinear growth is sufficient for the argument), and hence $T_e$ tends to $T_b$ at high temperatures even in the presence of finite $P_0$. We establish a \emph{temperature calibration} by associating the value $M(T_e)$ of the measurand with the temperature $T_b$.

Continuing the experiment, one obtains readings of the secondary electron thermometer by measuring $M$ and inverting the temperature calibration relation. It is evident that $T_{meas}$ obtained in this manner differs from the true $T_e$. Within our model, they are related by
\begin{equation}
    h(T_{meas}) = h(T_e) - P_0.
\end{equation}
To finalize our analysis, we rewrite the heat balance law Eq.~(\ref{eq:e}) in terms of $T_{meas}$ instead of $T_e$. We find
\begin{eqnarray}
P_{ext} & = & h(T_{meas}) - h(T_b).\label{eq:Qlaw1}
\end{eqnarray}
In what appears to be a fortuitous coincidence, the background heating term has been eliminated. 

\subsection{Three thermal bodies}

The thermal model for the full experiment presented in this work includes the two electron reservoirs and the phonon system as independent thermal bodies. We write
\begin{eqnarray}
    P_{ext}      + P_{0,e1} & = & h_{e1}(T_{e1}) - h_{e1}(T_p)\label{eq:e1}\\
                 P_{0,e2} & = & h_{e2}(T_{e2}) - h_{e2}(T_p)\label{eq:e2}\\
                 P_{0,p}  & = & \left[h_{e1}(T_{p}) - h_{e1}(T_{e1})\right] +\label{eq:p}\\ 
     & & \left[h_{e2}(T_{p}) - h_{e2}(T_{e2})\right] +\nonumber\\
     & & h_{p}(T_{p}) - h_{p}(T_{b}).\nonumber\\
     M_{1} & = & M_{1}(T_{e1}).\\
     M_{2} & = & M_{2}(T_{e2}).
\end{eqnarray}
Above, subscripts $e1$ and $e2$ refer to the two electron reservoirs, and $p$ to the phonon system. We allow each of the three heat links to be described by a different power law $h_i(T) = A_i T^{n_i}$. The experiment does not allow direct phonon heating or phonon thermometry. The two mesoscopic thermometers $M_1$ and $M_2$ sense the local electronic temperatures $T_{e1}$ and $T_{e2}$. To determine the electron-phonon coupling in reservoir $e1$, it is sufficient to apply electron heating is only to $e1$, as we will demonstrate shortly. Repeating the previous analysis, we model the full experimental protocol in two steps.

First, to model temperature calibration, we zero $P_{ext}$ and find the self-consistent solution for $(T_{e1}, T_{e2}, T_{p})$ as a function of $T_{b}$. These triplets should be interpreted as follows: After the calibration for the two independent electron thermometers has been carried out, the electron thermometer for reservoir $e1$ ($e2$) indicates temperature $T_b$ when the true physical temperature is $T_{e1}$ ($T_{e2}$). In the following, we will denote this temperature reading by $T_{meas,e1(e2)}$. Note that the apparent temperatures are always lower than the physical temperature if any of the $P_0$ terms is finite.

Second, to model the phonon heating step, we sweep $P_{ext}$ at a fixed $T_b$, and solve for self-consistent $(T_{e1}, T_{e2}, T_{p})$. Utilizing the temperature calibration model derived above, the simulated heater sweep experiment consists of the triplets $(P_{ext}, T_{meas,e1}, T_{meas,e2})$.

The preceding single-body analysis is directly applicable to the extraction of the phononic power law $h_p(T)$. To make the analogy exact, we note that Eqs.~(\ref{eq:e1})-(\ref{eq:p}) yield for the phononic thermal balance
\begin{equation}
    P_{ext} + P_{0,tot}= h_{p}(T_p) - h_{p}(T_b),\label{eq:pp}
\end{equation}
where we have introduced $P_{0,tot} = P_{0,e1} + P_{0,e2} + P_{0,P}$. Importantly, Eq.~(\ref{eq:pp}) is of the same form as Eq.~(\ref{eq:e}). Noting that $M_{2}$ is determined uniquely by $T_p$ (since $P_{0,e2}$ is fixed), it fulfills the condition for a non-linear secondary thermometer for $T_p$. This completes the analogy, implying that a power law $h_{p}(T)$ can be determined exactly by fitting a model
\begin{equation}
    P_{ext} = A_p\left(T_{meas,e2}\right)^{n_p} + B
\end{equation}
to the data.

For the analysis of the electron-phonon power law, Eqs.~(\ref{eq:e1})-(\ref{eq:p}) can be algebraically manipulated to yield the exact relation
\begin{eqnarray}
    P_{ext} = A_{e1} \left\{ \left[ \left(T_{meas,e1}\right)^{n_p} + C \right]^{n_e / n_p} \right.\nonumber\\
    - \left.\left[ \left(T_{meas,e2}\right)^{n_p} + C \right]^{n_e / n_p} \right\},\label{eq:Qlaw2}
\end{eqnarray}
where the free parameter $C$ stands for $P_{0,tot} / A_p$. The phonon heat law exponent $n_p$ can be determined independently in the manner described above, and is therefore considered a known value.

\subsection{Experimental data analysis}

Experimental data with a single thermal body is analyzed by fitting it with a model
\begin{equation}
    P_{ext} = A \left(T_{meas}\right)^n + B.
\end{equation}
In light of the preceding analysis, assuming the true $h(T)$ is a power law, it is accurately reproduced by the first term of the fitted model, and the constant term $B$ equals $-A T_b^n$.

For the three-body case, we find that the theoretically exact algebraic expression (\ref{eq:Qlaw2}) leads to numerical instabilities when used to fit experimental data. Instead, we use a simpler phenomenological model
\begin{equation}
    P_{ext} = A [\left(T_{meas,1}\right)^n - \left(T_{meas,2}\right)^n] + B,
\end{equation}
and verify the validity of the fit by observing small residuals.

In conclusion, our analytical studies show that thermal power laws can be determined from mesoscopic experiments using secondary thermometers even in the presence of unknown background heating offsets. We have verified the above analytical conclusions with numerical simulations. However, experimental statistical noise can introduce systematic biases when parameters of non-linear models are fitted.

\FloatBarrier

\bibliography{Thermometry}

\begin{thebibliography}{32}%
\makeatletter
\providecommand \@ifxundefined [1]{%
 \@ifx{#1\undefined}
}%
\providecommand \@ifnum [1]{%
 \ifnum #1\expandafter \@firstoftwo
 \else \expandafter \@secondoftwo
 \fi
}%
\providecommand \@ifx [1]{%
 \ifx #1\expandafter \@firstoftwo
 \else \expandafter \@secondoftwo
 \fi
}%
\providecommand \natexlab [1]{#1}%
\providecommand \enquote  [1]{``#1''}%
\providecommand \bibnamefont  [1]{#1}%
\providecommand \bibfnamefont [1]{#1}%
\providecommand \citenamefont [1]{#1}%
\providecommand \href@noop [0]{\@secondoftwo}%
\providecommand \href [0]{\begingroup \@sanitize@url \@href}%
\providecommand \@href[1]{\@@startlink{#1}\@@href}%
\providecommand \@@href[1]{\endgroup#1\@@endlink}%
\providecommand \@sanitize@url [0]{\catcode `\\12\catcode `\$12\catcode
  `\&12\catcode `\#12\catcode `\^12\catcode `\_12\catcode `\%12\relax}%
\providecommand \@@startlink[1]{}%
\providecommand \@@endlink[0]{}%
\providecommand \url  [0]{\begingroup\@sanitize@url \@url }%
\providecommand \@url [1]{\endgroup\@href {#1}{\urlprefix }}%
\providecommand \urlprefix  [0]{URL }%
\providecommand \Eprint [0]{\href }%
\providecommand \doibase [0]{http://dx.doi.org/}%
\providecommand \selectlanguage [0]{\@gobble}%
\providecommand \bibinfo  [0]{\@secondoftwo}%
\providecommand \bibfield  [0]{\@secondoftwo}%
\providecommand \translation [1]{[#1]}%
\providecommand \BibitemOpen [0]{}%
\providecommand \bibitemStop [0]{}%
\providecommand \bibitemNoStop [0]{.\EOS\space}%
\providecommand \EOS [0]{\spacefactor3000\relax}%
\providecommand \BibitemShut  [1]{\csname bibitem#1\endcsname}%
\let\auto@bib@innerbib\@empty
\bibitem [{\citenamefont {Giazotto}\ \emph {et~al.}(2006)\citenamefont
  {Giazotto}, \citenamefont {Heikkil{\"{a}}}, \citenamefont {Luukanen},
  \citenamefont {Savin},\ and\ \citenamefont {Pekola}}]{Giazotto2006}%
  \BibitemOpen
  \bibfield  {author} {\bibinfo {author} {\bibfnamefont {F.}~\bibnamefont
  {Giazotto}}, \bibinfo {author} {\bibfnamefont {T.~T.}\ \bibnamefont
  {Heikkil{\"{a}}}}, \bibinfo {author} {\bibfnamefont {A.}~\bibnamefont
  {Luukanen}}, \bibinfo {author} {\bibfnamefont {A.~M.}\ \bibnamefont {Savin}},
  \ and\ \bibinfo {author} {\bibfnamefont {J.~P.}\ \bibnamefont {Pekola}},\
  }\bibfield  {title} {\enquote {\bibinfo {title} {{Opportunities for
  mesoscopics in thermometry and refrigeration: Physics and applications}},}\
  }\href {http://link.aps.org/abstract/RMP/v78/p217} {\bibfield  {journal}
  {\bibinfo  {journal} {Rev. Mod. Phys.}\ }\textbf {\bibinfo {volume} {78}},\
  \bibinfo {pages} {217--258} (\bibinfo {year} {2006})}\BibitemShut {NoStop}%
\bibitem [{\citenamefont {Irwin}\ and\ \citenamefont
  {Hilton}(2005)}]{Irwin2005}%
  \BibitemOpen
  \bibfield  {author} {\bibinfo {author} {\bibfnamefont {K.}~\bibnamefont
  {Irwin}}\ and\ \bibinfo {author} {\bibfnamefont {G.}~\bibnamefont {Hilton}},\
  }\enquote {\bibinfo {title} {Transition-edge sensors},}\ in\ \href {\doibase
  10.1007/10933596_3} {\emph {\bibinfo {booktitle} {Cryogenic Particle
  Detection}}},\ \bibinfo {editor} {edited by\ \bibinfo {editor} {\bibfnamefont
  {C.}~\bibnamefont {Enss}}}\ (\bibinfo  {publisher} {Springer Berlin
  Heidelberg},\ \bibinfo {address} {Berlin, Heidelberg},\ \bibinfo {year}
  {2005})\ pp.\ \bibinfo {pages} {63--150}\BibitemShut {NoStop}%
\bibitem [{\citenamefont {Banerjee}\ \emph {et~al.}(2017)\citenamefont
  {Banerjee}, \citenamefont {Heiblum}, \citenamefont {Rosenblatt},
  \citenamefont {Oreg}, \citenamefont {Feldman}, \citenamefont {Stern},\ and\
  \citenamefont {Umansky}}]{Banerjee2017}%
  \BibitemOpen
  \bibfield  {author} {\bibinfo {author} {\bibfnamefont {M.}~\bibnamefont
  {Banerjee}}, \bibinfo {author} {\bibfnamefont {M.}~\bibnamefont {Heiblum}},
  \bibinfo {author} {\bibfnamefont {A.}~\bibnamefont {Rosenblatt}}, \bibinfo
  {author} {\bibfnamefont {Y.}~\bibnamefont {Oreg}}, \bibinfo {author}
  {\bibfnamefont {D.~E.}\ \bibnamefont {Feldman}}, \bibinfo {author}
  {\bibfnamefont {A.}~\bibnamefont {Stern}}, \ and\ \bibinfo {author}
  {\bibfnamefont {V.}~\bibnamefont {Umansky}},\ }\bibfield  {title} {\enquote
  {\bibinfo {title} {{Observed quantization of anyonic heat flow}},}\ }\href
  {\doibase 10.1038/nature22052} {\bibfield  {journal} {\bibinfo  {journal}
  {Nature}\ }\textbf {\bibinfo {volume} {545}},\ \bibinfo {pages} {75--79}
  (\bibinfo {year} {2017})}\BibitemShut {NoStop}%
\bibitem [{\citenamefont {Schwab}\ \emph {et~al.}(2000)\citenamefont {Schwab},
  \citenamefont {Henriksen}, \citenamefont {Worlock},\ and\ \citenamefont
  {Roukes}}]{Schwab2000}%
  \BibitemOpen
  \bibfield  {author} {\bibinfo {author} {\bibfnamefont {K.}~\bibnamefont
  {Schwab}}, \bibinfo {author} {\bibfnamefont {E.~A.}\ \bibnamefont
  {Henriksen}}, \bibinfo {author} {\bibfnamefont {J.~M.}\ \bibnamefont
  {Worlock}}, \ and\ \bibinfo {author} {\bibfnamefont {M.~L.}\ \bibnamefont
  {Roukes}},\ }\bibfield  {title} {\enquote {\bibinfo {title} {{Measurement of
  the quantum of thermal conductance}},}\ }\href
  {http://dx.doi.org/10.1038/35010065} {\bibfield  {journal} {\bibinfo
  {journal} {Nature}\ }\textbf {\bibinfo {volume} {404}},\ \bibinfo {pages}
  {974--977} (\bibinfo {year} {2000})}\BibitemShut {NoStop}%
\bibitem [{\citenamefont {L{\'{o}}pez}\ \emph {et~al.}(2014)\citenamefont
  {L{\'{o}}pez}, \citenamefont {Lee}, \citenamefont {Serra},\ and\
  \citenamefont {Lim}}]{Lopez2014}%
  \BibitemOpen
  \bibfield  {author} {\bibinfo {author} {\bibfnamefont {R.}~\bibnamefont
  {L{\'{o}}pez}}, \bibinfo {author} {\bibfnamefont {M.}~\bibnamefont {Lee}},
  \bibinfo {author} {\bibfnamefont {L.}~\bibnamefont {Serra}}, \ and\ \bibinfo
  {author} {\bibfnamefont {J.~S.}\ \bibnamefont {Lim}},\ }\bibfield  {title}
  {\enquote {\bibinfo {title} {{Thermoelectrical detection of Majorana
  states}},}\ }\href {\doibase 10.1103/PhysRevB.89.205418} {\bibfield
  {journal} {\bibinfo  {journal} {Phys. Rev. B}\ }\textbf {\bibinfo {volume}
  {89}},\ \bibinfo {pages} {205418} (\bibinfo {year} {2014})}\BibitemShut
  {NoStop}%
\bibitem [{\citenamefont {Ramos-Andrade}\ \emph {et~al.}(2016)\citenamefont
  {Ramos-Andrade}, \citenamefont {{\'{A}}valos-Ovando}, \citenamefont
  {Orellana},\ and\ \citenamefont {Ulloa}}]{Ramos-Andrade2016}%
  \BibitemOpen
  \bibfield  {author} {\bibinfo {author} {\bibfnamefont {J.~P.}\ \bibnamefont
  {Ramos-Andrade}}, \bibinfo {author} {\bibfnamefont {O.}~\bibnamefont
  {{\'{A}}valos-Ovando}}, \bibinfo {author} {\bibfnamefont {P.~A.}\
  \bibnamefont {Orellana}}, \ and\ \bibinfo {author} {\bibfnamefont {S.~E.}\
  \bibnamefont {Ulloa}},\ }\bibfield  {title} {\enquote {\bibinfo {title}
  {{Thermoelectric transport through Majorana bound states and violation of
  Wiedemann-Franz law}},}\ }\href {\doibase 10.1103/PhysRevB.94.155436}
  {\bibfield  {journal} {\bibinfo  {journal} {Phys. Rev. B}\ }\textbf {\bibinfo
  {volume} {94}},\ \bibinfo {pages} {155436} (\bibinfo {year}
  {2016})}\BibitemShut {NoStop}%
\bibitem [{\citenamefont {Molignini}, \citenamefont {van Nieuwenburg},\ and\
  \citenamefont {Chitra}(2017)}]{Molignini2017}%
  \BibitemOpen
  \bibfield  {author} {\bibinfo {author} {\bibfnamefont {P.}~\bibnamefont
  {Molignini}}, \bibinfo {author} {\bibfnamefont {E.}~\bibnamefont {van
  Nieuwenburg}}, \ and\ \bibinfo {author} {\bibfnamefont {R.}~\bibnamefont
  {Chitra}},\ }\bibfield  {title} {\enquote {\bibinfo {title} {{Sensing
  Floquet-Majorana fermions via heat transfer}},}\ }\href {\doibase
  10.1103/PhysRevB.96.125144} {\bibfield  {journal} {\bibinfo  {journal} {Phys.
  Rev. B}\ }\textbf {\bibinfo {volume} {96}},\ \bibinfo {pages} {125144}
  (\bibinfo {year} {2017})}\BibitemShut {NoStop}%
\bibitem [{\citenamefont {Dubos}\ \emph {et~al.}(2001)\citenamefont {Dubos},
  \citenamefont {Courtois}, \citenamefont {Pannetier}, \citenamefont {Wilhelm},
  \citenamefont {Zaikin},\ and\ \citenamefont {Sch{\"{o}}n}}]{Dubos2001}%
  \BibitemOpen
  \bibfield  {author} {\bibinfo {author} {\bibfnamefont {P.}~\bibnamefont
  {Dubos}}, \bibinfo {author} {\bibfnamefont {H.}~\bibnamefont {Courtois}},
  \bibinfo {author} {\bibfnamefont {B.}~\bibnamefont {Pannetier}}, \bibinfo
  {author} {\bibfnamefont {F.~K.}\ \bibnamefont {Wilhelm}}, \bibinfo {author}
  {\bibfnamefont {A.~D.}\ \bibnamefont {Zaikin}}, \ and\ \bibinfo {author}
  {\bibfnamefont {G.}~\bibnamefont {Sch{\"{o}}n}},\ }\bibfield  {title}
  {\enquote {\bibinfo {title} {{Josephson critical current in a long mesoscopic
  S-N-S junction}},}\ }\href {\doibase 10.1103/PhysRevB.63.064502} {\bibfield
  {journal} {\bibinfo  {journal} {Phys. Rev. B}\ }\textbf {\bibinfo {volume}
  {63}},\ \bibinfo {pages} {064502} (\bibinfo {year} {2001})}\BibitemShut
  {NoStop}%
\bibitem [{\citenamefont {Courtois}\ \emph {et~al.}(2008)\citenamefont
  {Courtois}, \citenamefont {Meschke}, \citenamefont {Peltonen},\ and\
  \citenamefont {Pekola}}]{Courtois2008}%
  \BibitemOpen
  \bibfield  {author} {\bibinfo {author} {\bibfnamefont {H.}~\bibnamefont
  {Courtois}}, \bibinfo {author} {\bibfnamefont {M.}~\bibnamefont {Meschke}},
  \bibinfo {author} {\bibfnamefont {J.~T.}\ \bibnamefont {Peltonen}}, \ and\
  \bibinfo {author} {\bibfnamefont {J.~P.}\ \bibnamefont {Pekola}},\ }\bibfield
   {title} {\enquote {\bibinfo {title} {{Origin of Hysteresis in a Proximity
  Josephson Junction}},}\ }\href {\doibase 10.1103/PhysRevLett.101.067002}
  {\bibfield  {journal} {\bibinfo  {journal} {Phys. Rev. Lett.}\ }\textbf
  {\bibinfo {volume} {101}},\ \bibinfo {pages} {067002} (\bibinfo {year}
  {2008})}\BibitemShut {NoStop}%
\bibitem [{\citenamefont {Govenius}\ \emph {et~al.}(2014)\citenamefont
  {Govenius}, \citenamefont {Lake}, \citenamefont {Tan}, \citenamefont
  {Pietil{\"{a}}}, \citenamefont {Julin}, \citenamefont {Maasilta},
  \citenamefont {Virtanen},\ and\ \citenamefont
  {M{\"{o}}tt{\"{o}}nen}}]{Govenius2014}%
  \BibitemOpen
  \bibfield  {author} {\bibinfo {author} {\bibfnamefont {J.}~\bibnamefont
  {Govenius}}, \bibinfo {author} {\bibfnamefont {R.~E.}\ \bibnamefont {Lake}},
  \bibinfo {author} {\bibfnamefont {K.~Y.}\ \bibnamefont {Tan}}, \bibinfo
  {author} {\bibfnamefont {V.}~\bibnamefont {Pietil{\"{a}}}}, \bibinfo {author}
  {\bibfnamefont {J.~K.}\ \bibnamefont {Julin}}, \bibinfo {author}
  {\bibfnamefont {I.~J.}\ \bibnamefont {Maasilta}}, \bibinfo {author}
  {\bibfnamefont {P.}~\bibnamefont {Virtanen}}, \ and\ \bibinfo {author}
  {\bibfnamefont {M.}~\bibnamefont {M{\"{o}}tt{\"{o}}nen}},\ }\bibfield
  {title} {\enquote {\bibinfo {title} {{Microwave nanobolometer based on
  proximity Josephson junctions}},}\ }\href {\doibase
  10.1103/PhysRevB.90.064505} {\bibfield  {journal} {\bibinfo  {journal} {Phys.
  Rev. B}\ }\textbf {\bibinfo {volume} {90}},\ \bibinfo {pages} {064505}
  (\bibinfo {year} {2014})}\BibitemShut {NoStop}%
\bibitem [{\citenamefont {Wang}, \citenamefont {Saira},\ and\ \citenamefont
  {Pekola}(2018)}]{Wang2018}%
  \BibitemOpen
  \bibfield  {author} {\bibinfo {author} {\bibfnamefont {L.~B.}\ \bibnamefont
  {Wang}}, \bibinfo {author} {\bibfnamefont {O.-P.}\ \bibnamefont {Saira}}, \
  and\ \bibinfo {author} {\bibfnamefont {J.~P.}\ \bibnamefont {Pekola}},\
  }\bibfield  {title} {\enquote {\bibinfo {title} {{Fast thermometry with a
  proximity Josephson junction}},}\ }\href {\doibase 10.1063/1.5010236}
  {\bibfield  {journal} {\bibinfo  {journal} {Appl. Phys. Lett.}\ }\textbf
  {\bibinfo {volume} {112}},\ \bibinfo {pages} {013105} (\bibinfo {year}
  {2018})}\BibitemShut {NoStop}%
\bibitem [{\citenamefont {Anghel}, \citenamefont {Caraiani},\ and\
  \citenamefont {Galperin}(2018)}]{Anghel2018}%
  \BibitemOpen
  \bibfield  {author} {\bibinfo {author} {\bibfnamefont {D.-V.}\ \bibnamefont
  {Anghel}}, \bibinfo {author} {\bibfnamefont {C.}~\bibnamefont {Caraiani}}, \
  and\ \bibinfo {author} {\bibfnamefont {Y.~M.}\ \bibnamefont {Galperin}},\
  }\bibfield  {title} {\enquote {\bibinfo {title} {{Crossover in the
  electron-phonon heat exchange in layered nanostructures}},}\ }\href
  {http://arxiv.org/abs/1810.02360} {\  (\bibinfo {year} {2018})},\ \Eprint
  {http://arxiv.org/abs/1810.02360} {arXiv:1810.02360} \BibitemShut {NoStop}%
\bibitem [{\citenamefont {Karvonen}\ and\ \citenamefont
  {Maasilta}(2007{\natexlab{a}})}]{Karvonen2007}%
  \BibitemOpen
  \bibfield  {author} {\bibinfo {author} {\bibfnamefont {J.~T.}\ \bibnamefont
  {Karvonen}}\ and\ \bibinfo {author} {\bibfnamefont {I.~J.}\ \bibnamefont
  {Maasilta}},\ }\bibfield  {title} {\enquote {\bibinfo {title} {{Influence of
  Phonon Dimensionality on Electron Energy Relaxation}},}\ }\href {\doibase
  10.1103/PhysRevLett.99.145503} {\bibfield  {journal} {\bibinfo  {journal}
  {Phys. Rev. Lett.}\ }\textbf {\bibinfo {volume} {99}},\ \bibinfo {pages}
  {145503} (\bibinfo {year} {2007}{\natexlab{a}})}\BibitemShut {NoStop}%
\bibitem [{\citenamefont {Karvonen}\ and\ \citenamefont
  {Maasilta}(2007{\natexlab{b}})}]{Karvonen2007b}%
  \BibitemOpen
  \bibfield  {author} {\bibinfo {author} {\bibfnamefont {J.~T.}\ \bibnamefont
  {Karvonen}}\ and\ \bibinfo {author} {\bibfnamefont {I.~J.}\ \bibnamefont
  {Maasilta}},\ }\bibfield  {title} {\enquote {\bibinfo {title} {{Observation
  of phonon dimensionality effects on electron energy relaxation}},}\ }\href
  {\doibase 10.1088/1742-6596/92/1/012043} {\bibfield  {journal} {\bibinfo
  {journal} {J. Phys.: Conf. Ser}\ }\textbf {\bibinfo {volume} {92}},\ \bibinfo
  {pages} {012043} (\bibinfo {year} {2007}{\natexlab{b}})}\BibitemShut
  {NoStop}%
\bibitem [{\citenamefont {Shepherd}(1972)}]{Shepherd1972}%
  \BibitemOpen
  \bibfield  {author} {\bibinfo {author} {\bibfnamefont {J.~G.}\ \bibnamefont
  {Shepherd}},\ }\bibfield  {title} {\enquote {\bibinfo {title} {{Supercurrents
  Through Thick, Clean S-N-S Sandwiches}},}\ }\href {\doibase
  10.1098/rspa.1972.0018} {\bibfield  {journal} {\bibinfo  {journal} {Proc. R.
  Soc. Lond. A}\ }\textbf {\bibinfo {volume} {326}},\ \bibinfo {pages}
  {421--430} (\bibinfo {year} {1972})}\BibitemShut {NoStop}%
\bibitem [{\citenamefont {Clarke}(1969)}]{Clarke1969}%
  \BibitemOpen
  \bibfield  {author} {\bibinfo {author} {\bibfnamefont {J.}~\bibnamefont
  {Clarke}},\ }\bibfield  {title} {\enquote {\bibinfo {title} {{Supercurrents
  in lead—copper—-lead sandwiches}},}\ }\href {\doibase
  10.1098/rspa.1969.0020} {\bibfield  {journal} {\bibinfo  {journal} {Proc. R.
  Soc. Lond. A}\ }\textbf {\bibinfo {volume} {308}},\ \bibinfo {pages}
  {447--471} (\bibinfo {year} {1969})}\BibitemShut {NoStop}%
\bibitem [{\citenamefont {Stewart}(1968)}]{Stewart1968}%
  \BibitemOpen
  \bibfield  {author} {\bibinfo {author} {\bibfnamefont {W.~C.}\ \bibnamefont
  {Stewart}},\ }\bibfield  {title} {\enquote {\bibinfo {title}
  {{Current-voltage characteristics of Josephson junctions}},}\ }\href
  {\doibase 10.1063/1.1651991} {\bibfield  {journal} {\bibinfo  {journal}
  {Appl. Phys. Lett}\ }\textbf {\bibinfo {volume} {12}},\ \bibinfo {pages}
  {277} (\bibinfo {year} {1968})}\BibitemShut {NoStop}%
\bibitem [{\citenamefont {McCumber}(1968)}]{McCumber1968}%
  \BibitemOpen
  \bibfield  {author} {\bibinfo {author} {\bibfnamefont {D.~E.}\ \bibnamefont
  {McCumber}},\ }\bibfield  {title} {\enquote {\bibinfo {title} {{Effect of ac
  impedance on dc voltage-current characteristics of superconductor weak-link
  junctions}},}\ }\href {\doibase 10.1063/1.1656743} {\bibfield  {journal}
  {\bibinfo  {journal} {J Appl. Phys.}\ }\textbf {\bibinfo {volume} {39}},\
  \bibinfo {pages} {3113--3118} (\bibinfo {year} {1968})}\BibitemShut {NoStop}%
\bibitem [{\citenamefont {Likharev}(1979)}]{Likharev1979}%
  \BibitemOpen
  \bibfield  {author} {\bibinfo {author} {\bibfnamefont {K.~K.}\ \bibnamefont
  {Likharev}},\ }\bibfield  {title} {\enquote {\bibinfo {title}
  {{Superconducting weak links}},}\ }\href {\doibase 10.1103/RevModPhys.51.101}
  {\bibfield  {journal} {\bibinfo  {journal} {Rev. Mod. Phys.}\ }\textbf
  {\bibinfo {volume} {51}},\ \bibinfo {pages} {101--159} (\bibinfo {year}
  {1979})}\BibitemShut {NoStop}%
\bibitem [{\citenamefont {Jabdaraghi}\ \emph {et~al.}(2016)\citenamefont
  {Jabdaraghi}, \citenamefont {Peltonen}, \citenamefont {Saira},\ and\
  \citenamefont {Pekola}}]{Jabdaraghi2016}%
  \BibitemOpen
  \bibfield  {author} {\bibinfo {author} {\bibfnamefont {R.~N.}\ \bibnamefont
  {Jabdaraghi}}, \bibinfo {author} {\bibfnamefont {J.~T.}\ \bibnamefont
  {Peltonen}}, \bibinfo {author} {\bibfnamefont {O.-P.}\ \bibnamefont {Saira}},
  \ and\ \bibinfo {author} {\bibfnamefont {J.~P.}\ \bibnamefont {Pekola}},\
  }\bibfield  {title} {\enquote {\bibinfo {title} {{Low-temperature
  characterization of Nb-Cu-Nb weak links with Ar ion-cleaned interfaces}},}\
  }\href {\doibase 10.1063/1.4940979} {\bibfield  {journal} {\bibinfo
  {journal} {Appl. Phys. Lett.}\ }\textbf {\bibinfo {volume} {108}},\ \bibinfo
  {pages} {042604} (\bibinfo {year} {2016})}\BibitemShut {NoStop}%
\bibitem [{\citenamefont {Zgirski}\ \emph {et~al.}(2018)\citenamefont
  {Zgirski}, \citenamefont {Foltyn}, \citenamefont {Savin}, \citenamefont
  {Norowski}, \citenamefont {Meschke},\ and\ \citenamefont
  {Pekola}}]{Zgirski2018}%
  \BibitemOpen
  \bibfield  {author} {\bibinfo {author} {\bibfnamefont {M.}~\bibnamefont
  {Zgirski}}, \bibinfo {author} {\bibfnamefont {M.}~\bibnamefont {Foltyn}},
  \bibinfo {author} {\bibfnamefont {A.}~\bibnamefont {Savin}}, \bibinfo
  {author} {\bibfnamefont {K.}~\bibnamefont {Norowski}}, \bibinfo {author}
  {\bibfnamefont {M.}~\bibnamefont {Meschke}}, \ and\ \bibinfo {author}
  {\bibfnamefont {J.}~\bibnamefont {Pekola}},\ }\bibfield  {title} {\enquote
  {\bibinfo {title} {{Nanosecond Thermometry with Josephson Junctions}},}\
  }\href {\doibase 10.1103/PhysRevApplied.10.044068} {\bibfield  {journal}
  {\bibinfo  {journal} {Phys. Rev. Applied}\ }\textbf {\bibinfo {volume}
  {10}},\ \bibinfo {pages} {044068} (\bibinfo {year} {2018})}\BibitemShut
  {NoStop}%
\bibitem [{\citenamefont {Roukes}\ \emph {et~al.}(1985)\citenamefont {Roukes},
  \citenamefont {Freeman}, \citenamefont {Germain}, \citenamefont
  {Richardson},\ and\ \citenamefont {Ketchen}}]{Roukes1985}%
  \BibitemOpen
  \bibfield  {author} {\bibinfo {author} {\bibfnamefont {M.~L.}\ \bibnamefont
  {Roukes}}, \bibinfo {author} {\bibfnamefont {M.~R.}\ \bibnamefont {Freeman}},
  \bibinfo {author} {\bibfnamefont {R.~S.}\ \bibnamefont {Germain}}, \bibinfo
  {author} {\bibfnamefont {R.~C.}\ \bibnamefont {Richardson}}, \ and\ \bibinfo
  {author} {\bibfnamefont {M.~B.}\ \bibnamefont {Ketchen}},\ }\bibfield
  {title} {\enquote {\bibinfo {title} {{Hot electrons and energy transport in
  metals at millikelvin temperatures}},}\ }\href {\doibase
  10.1103/PhysRevLett.55.422} {\bibfield  {journal} {\bibinfo  {journal} {Phys.
  Rev. Lett.}\ }\textbf {\bibinfo {volume} {55}},\ \bibinfo {pages} {422--425}
  (\bibinfo {year} {1985})}\BibitemShut {NoStop}%
\bibitem [{\citenamefont {Wellstood}, \citenamefont {Urbina},\ and\
  \citenamefont {Clarke}(1994)}]{Wellstood1994}%
  \BibitemOpen
  \bibfield  {author} {\bibinfo {author} {\bibfnamefont {F.~C.}\ \bibnamefont
  {Wellstood}}, \bibinfo {author} {\bibfnamefont {C.}~\bibnamefont {Urbina}}, \
  and\ \bibinfo {author} {\bibfnamefont {J.}~\bibnamefont {Clarke}},\
  }\bibfield  {title} {\enquote {\bibinfo {title} {{Hot-electron effects in
  metals}},}\ }\href {\doibase 10.1103/PhysRevB.49.5942} {\bibfield  {journal}
  {\bibinfo  {journal} {Phys. Rev. B}\ }\textbf {\bibinfo {volume} {49}},\
  \bibinfo {pages} {5942--5955} (\bibinfo {year} {1994})}\BibitemShut {NoStop}%
\bibitem [{\citenamefont {Qu}, \citenamefont {Cleland},\ and\ \citenamefont
  {Geller}(2005)}]{Qu2005}%
  \BibitemOpen
  \bibfield  {author} {\bibinfo {author} {\bibfnamefont {S.-X.}\ \bibnamefont
  {Qu}}, \bibinfo {author} {\bibfnamefont {A.~N.}\ \bibnamefont {Cleland}}, \
  and\ \bibinfo {author} {\bibfnamefont {M.~R.}\ \bibnamefont {Geller}},\
  }\bibfield  {title} {\enquote {\bibinfo {title} {{Hot electrons in
  low-dimensional phonon systems}},}\ }\href {\doibase
  10.1103/PhysRevB.72.224301} {\bibfield  {journal} {\bibinfo  {journal} {Phys.
  Rev. B}\ }\textbf {\bibinfo {volume} {72}},\ \bibinfo {pages} {224301}
  (\bibinfo {year} {2005})}\BibitemShut {NoStop}%
\bibitem [{\citenamefont {Echternach}\ \emph {et~al.}(1992)\citenamefont
  {Echternach}, \citenamefont {Thoman}, \citenamefont {Gould},\ and\
  \citenamefont {Bozler}}]{Echternach1992}%
  \BibitemOpen
  \bibfield  {author} {\bibinfo {author} {\bibfnamefont {P.~M.}\ \bibnamefont
  {Echternach}}, \bibinfo {author} {\bibfnamefont {M.~R.}\ \bibnamefont
  {Thoman}}, \bibinfo {author} {\bibfnamefont {C.~M.}\ \bibnamefont {Gould}}, \
  and\ \bibinfo {author} {\bibfnamefont {H.~M.}\ \bibnamefont {Bozler}},\
  }\bibfield  {title} {\enquote {\bibinfo {title} {{Electron-phonon scattering
  rates in disordered metallic films below 1 K}},}\ }\href {\doibase
  10.1103/PhysRevB.46.10339} {\bibfield  {journal} {\bibinfo  {journal} {Phys.
  Rev. B}\ }\textbf {\bibinfo {volume} {46}},\ \bibinfo {pages} {10339--10344}
  (\bibinfo {year} {1992})}\BibitemShut {NoStop}%
\bibitem [{\citenamefont {Anghel}\ \emph {et~al.}(1998)\citenamefont {Anghel},
  \citenamefont {Pekola}, \citenamefont {Leivo}, \citenamefont {Suoknuuti},\
  and\ \citenamefont {Manninen}}]{Anghel1998}%
  \BibitemOpen
  \bibfield  {author} {\bibinfo {author} {\bibfnamefont {D.~V.}\ \bibnamefont
  {Anghel}}, \bibinfo {author} {\bibfnamefont {J.~P.}\ \bibnamefont {Pekola}},
  \bibinfo {author} {\bibfnamefont {M.~M.}\ \bibnamefont {Leivo}}, \bibinfo
  {author} {\bibfnamefont {J.~K.}\ \bibnamefont {Suoknuuti}}, \ and\ \bibinfo
  {author} {\bibfnamefont {M.}~\bibnamefont {Manninen}},\ }\bibfield  {title}
  {\enquote {\bibinfo {title} {{Properties of the Phonon Gas in Ultrathin
  Membranes at Low Temperature}},}\ }\href {\doibase
  10.1103/PhysRevLett.81.2958} {\bibfield  {journal} {\bibinfo  {journal}
  {Phys. Rev. Lett.}\ }\textbf {\bibinfo {volume} {81}},\ \bibinfo {pages}
  {2958--2961} (\bibinfo {year} {1998})}\BibitemShut {NoStop}%
\bibitem [{\citenamefont {K{\"{u}}hn}\ \emph {et~al.}(2004)\citenamefont
  {K{\"{u}}hn}, \citenamefont {Anghel}, \citenamefont {Pekola}, \citenamefont
  {Manninen},\ and\ \citenamefont {Galperin}}]{Kuhn2004}%
  \BibitemOpen
  \bibfield  {author} {\bibinfo {author} {\bibfnamefont {T.}~\bibnamefont
  {K{\"{u}}hn}}, \bibinfo {author} {\bibfnamefont {D.~V.}\ \bibnamefont
  {Anghel}}, \bibinfo {author} {\bibfnamefont {J.~P.}\ \bibnamefont {Pekola}},
  \bibinfo {author} {\bibfnamefont {M.}~\bibnamefont {Manninen}}, \ and\
  \bibinfo {author} {\bibfnamefont {Y.~M.}\ \bibnamefont {Galperin}},\
  }\bibfield  {title} {\enquote {\bibinfo {title} {{Heat transport in ultrathin
  dielectric membranes and bridges}},}\ }\href {\doibase
  10.1103/PhysRevB.70.125425} {\bibfield  {journal} {\bibinfo  {journal} {Phys.
  Rev. B}\ }\textbf {\bibinfo {volume} {70}},\ \bibinfo {pages} {125425}
  (\bibinfo {year} {2004})}\BibitemShut {NoStop}%
\bibitem [{\citenamefont {Cojocaru}\ and\ \citenamefont
  {Anghel}(2016)}]{Cojocaru2016}%
  \BibitemOpen
  \bibfield  {author} {\bibinfo {author} {\bibfnamefont {S.}~\bibnamefont
  {Cojocaru}}\ and\ \bibinfo {author} {\bibfnamefont {D.~V.}\ \bibnamefont
  {Anghel}},\ }\bibfield  {title} {\enquote {\bibinfo {title} {{Low-temperature
  electron-phonon heat transfer in metal films}},}\ }\href {\doibase
  10.1103/PhysRevB.93.115405} {\bibfield  {journal} {\bibinfo  {journal} {Phys.
  Rev. B}\ }\textbf {\bibinfo {volume} {93}},\ \bibinfo {pages} {115405}
  (\bibinfo {year} {2016})}\BibitemShut {NoStop}%
\bibitem [{\citenamefont {Anghel}\ and\ \citenamefont
  {Cojocaru}(2017)}]{Anghel2017}%
  \BibitemOpen
  \bibfield  {author} {\bibinfo {author} {\bibfnamefont {D.-V.}\ \bibnamefont
  {Anghel}}\ and\ \bibinfo {author} {\bibfnamefont {S.}~\bibnamefont
  {Cojocaru}},\ }\bibfield  {title} {\enquote {\bibinfo {title}
  {{Electron-phonon heat exchange in quasi-two-dimensional nanolayers}},}\
  }\href {\doibase 10.1140/epjb/e2017-80111-y} {\bibfield  {journal} {\bibinfo
  {journal} {Eur. Phys. J. B}\ }\textbf {\bibinfo {volume} {90}},\ \bibinfo
  {pages} {260} (\bibinfo {year} {2017})}\BibitemShut {NoStop}%
\bibitem [{\citenamefont {Lin}\ and\ \citenamefont
  {Pugacz-Muraszkiewicz}(1972)}]{Lin1972}%
  \BibitemOpen
  \bibfield  {author} {\bibinfo {author} {\bibfnamefont {S.~C.~H.}\
  \bibnamefont {Lin}}\ and\ \bibinfo {author} {\bibfnamefont {I.}~\bibnamefont
  {Pugacz-Muraszkiewicz}},\ }\bibfield  {title} {\enquote {\bibinfo {title}
  {{Local Stress Measurement in Thin Thermal SiO2 Films on Si Substrates}},}\
  }\href {\doibase 10.1063/1.1660794} {\bibfield  {journal} {\bibinfo
  {journal} {J. Appl. Phys.}\ }\textbf {\bibinfo {volume} {43}},\ \bibinfo
  {pages} {119} (\bibinfo {year} {1972})}\BibitemShut {NoStop}%
\bibitem [{\citenamefont {Klokholm}(1969)}]{Klokholm1969}%
  \BibitemOpen
  \bibfield  {author} {\bibinfo {author} {\bibfnamefont {E.}~\bibnamefont
  {Klokholm}},\ }\bibfield  {title} {\enquote {\bibinfo {title} {{Intrinsic
  Stress in Evaporated Metal Films}},}\ }\href {\doibase 10.1116/1.1492645}
  {\bibfield  {journal} {\bibinfo  {journal} {J. Vac. Sci. Tech.}\ }\textbf
  {\bibinfo {volume} {6}},\ \bibinfo {pages} {138} (\bibinfo {year}
  {1969})}\BibitemShut {NoStop}%
\bibitem [{\citenamefont {Corruccini}\ and\ \citenamefont
  {Gniewek}(1961)}]{Corruccini1961}%
  \BibitemOpen
  \bibfield  {author} {\bibinfo {author} {\bibfnamefont {R.~J.}\ \bibnamefont
  {Corruccini}}\ and\ \bibinfo {author} {\bibfnamefont {J.~J.}\ \bibnamefont
  {Gniewek}},\ }\href@noop {} {\enquote {\bibinfo {title} {{Thermal Expansion
  of Technical Solids at Low Temperatures}},}\ }\bibinfo {type} {Tech. Rep.}\
  (\bibinfo {year} {1961})\BibitemShut {NoStop}%
\end{thebibliography}%

\end{document}